\documentclass[
 prb,
 amsmath,amssymb,
 aps
]{revtex4-2}
\usepackage{graphicx}
\usepackage{appendix}
\usepackage{bm}
\usepackage{braket}
\makeatletter
\newsavebox{\@brx}
\begin{document}

\title{Metallic mean quasicrystals and their topological invariants}

\author{Anuradha Jagannathan}

\affiliation{Universit\'{e} Paris-Saclay, CNRS, Laboratoire de Physique des Solides, 91405 Orsay, France}

\date{}

\begin{abstract}
Topological invariants govern many important physical properties in condensed matter systems. In this work, we obtain the complete set of topological invariants for a family of one-dimensional quasicrystals. The first and best-studied member of the family is the Fibonacci chain, while the successive ones are known in the literature as silver, bronze... and  collectively as the metallic mean chains. By considering rational approximants, and by making use of the relationship between these chains and two dimensional Quantum Hall problems, we write down a gap labeling scheme for finite systems, and extend it to the quasiperiodic limit. We show, by numerical computations on open chains, that the proposed scheme correctly yields the winding numbers of edge states in each of the gaps, in all of the quasicrystals. We observe that their relationship to the 2D model leads, in the strict 1D chains family, to spectra forming a simplified Hofstadter ``butterfly" diagram, with the analogues of Landau levels appearing in the asymptotic limit. These topological invariants for noninteracting 1D chains can be expected to remain pertinent for superconducting systems where Majorana or Aharanov bound states occur, as shown in recent studies of interacting or Josephson coupled Fibonacci chains.
\end{abstract}

\maketitle
\section{Introduction}
Topological invariants, especially those which govern physical properties  are justifiably of great interest in condensed matter systems. In recent years, many studies have focused on the intrinsic topological properties of quasicrystals, and on their relation to higher dimensional periodic models. The connection between 2D quantum Hall (QH) problems and 1D quasicrystals have been pointed out in a number of early works of Bellissard and collaborators \cite{bellissardpapers}. More recently, on the theoretical front, connections between 2D and 1D problems have been discussed in a projected branes approach \cite{juricic} and via the Fibonacci-Hall model \cite{jaga2025}. The latter provides a convenient platform to confirm the ideas of topological equivalence between the quasicrystal and the Harper models that underlies experimental studies in optical waveguides \cite{kraus,zilberberg}. In other experiments, edge states and associated topological invariants have been observed in Fibonacci chains of polaritonic resonators \cite{tanese,baboux}. Our aim in this paper is to generalize results for topological invariants to all of the members of a class of 1D quasicrystals called the ``metallic mean" chains, thus providing a global description of topological invariants for an entire family of quasiperiodic structures. 

Given the importance of topological properties for applications, it is of interest to understand quasicrystalline systems in particular since they are less studied compared to crystalline systems.
This work gives a unified view of the topological invariants of an entire family of quasicrystals, and in doing so, marks a step forward in the understanding of quasiperiodic systems.  

Our approach defines a 2D QH problem for each of the finite periodic approximants of these quasicrystals. The complete set of 2D problems  corresponds to a subset of the Hofstadter butterfly diagram. These 2D models are used to write down a topological indexing ansatz for the family of quasiperiodic systems. To check the ansatz, we examine the winding of the edge states within each of the gaps, for finite open chains. The quasiperiodic limit is found as the size of the approximant is sent to infinity for all of the chains.

\section{The metallic mean quasicrystals and their approximants}
The Fibonacci quasicrystal has lesser-known relatives going by the names of the silver mean and bronze mean quasicrystals, and so on collectively called the 1D metallic mean quasicrystals. Some electronic properties of these 1D systems -- spectra, wavefunctions and wavepacket diffusion have been studied \cite{cerovski,thiem2009,thiem2011}. However, their topological characteristics have not so far been studied to our knowledge. Each of the metallic mean chains can be described in terms of an irrational number, $\omega_n$, the largest root of the equation $\omega^2=n\omega+1$. For $n=1$ this yields the golden mean, $ \omega_1=\frac{\sqrt{5}+1}{2}$, giving rise to the Fibonacci chain. The cases $n=2$ and $n=3$ correspond to the so-called silver and bronze means respectively.

For the $n$th member of the family, the infinite quasicrystal can be generated using substitution rules for the A-tile and the B-tile as follows:
\begin{eqnarray}
    B &\rightarrow& A \\ \nonumber
    A &\rightarrow& A^n B
    \label{eq:subst}
\end{eqnarray}
For example, in the case $n=2$, the substitution rules applied to an initial ``seed" B yield a series of finite chains as follows: $B \rightarrow A \rightarrow AAB \rightarrow AABAABA \rightarrow AABAABAAABAABAAAB \rightarrow ....$. 
The finite chains ${\mathcal{C}^{(k)}_n}$ are termed approximants.
The $n$th metallic mean quasicrystal is obtained from the series ${\mathcal{C}^{(k)}_n}$ in the limit $k\rightarrow \infty$.

The substitution matrix, $M$, relates the number of each type of tile before and after a substitution. Defining the vector $V^{(k)} = (N^{(k)}_B, N^{(k)}_A)$ where $N^{(k)}_{B(A)}$ are the number of B(A) tiles respectively, one has $V^{(k)} = M V^{(k-1)} $, with
\begin{eqnarray}
 M = \left(
 \begin{array}{cc}
0 & 1 \\
1 & n
 \end{array}
 \right)
\end{eqnarray}
The eigenvalues of $M$ are $\omega_n$ and $\omega^{-1}_n$, where
$\omega_n =(n + \sqrt{n^2+4})/2$. The Perron-Frobenius eigenvalue governs the rate of growth of the approximant chains as a function of $k$. That is, the number of tiles in the $k$th chain grows asymptotically as $\omega_n^k$. The corresponding eigenvector gives the ratio of B tiles and A tiles in the limit  $k\rightarrow \infty$. One finds $N^{(k)}_{B} :N^{(k)}_{A}$ is equal to $1:\omega_n$, in other words, the series of chains tends towards the periodic chain. The matrices $M$ describe Pisot systems -- that is, the modulus of one eigenvalue is greater than 1, the other less than 1. This property ensures that structure fluctuations are bounded in the infinite chain \cite{luck93} -- indeed, quasicrystals, like crystals, are hyperuniform structures \cite{torquato,baake2019}. This is not the case for substitutions such as $B\rightarrow A B\rightarrow B^n A$.
The eigenvalues have a continued fraction expansion given by $\omega_n = [n;n,n,n,...]$.
Truncating the continued fraction at the $k$th order yields rational approximants of $\omega_{n}$. One has 
\begin{eqnarray}
 \omega^{-1}_n = \lim_{k\rightarrow \infty} \omega^{-1}_{n,k}; \qquad \omega^{-1}_{n,k} = \frac{P^{(k-1)}_{n}}{P^{(k)}_{n}}
\end{eqnarray}
where the numbers $P^{(k)}_{n}$ satisfy the recursion relation
\begin{eqnarray}
    P^{(k)}_{n} = n P^{(k-1)}_{n} + P^{(k-2)}_{n} \qquad
    P^{(0)}_{n} = 0 ~; P^{(1)}_{n} = 1
    \label{eq:recur}
\end{eqnarray}
For $n=1$, the $P^{(k)}_{1}$ are the Fibonacci numbers. For $n=2$, the $P^{(k)}_{2}$, are known as the Pell numbers, and the ratio $P^{(k+1)}_{2}/P^{(k)}_{2}$ tends to the silver mean $\omega_2 = (\sqrt{2}+1)$ as $k\rightarrow \infty$. Similar relations hold for the other metallic means. 

It will be convenient to introduce another set of numbers, $Q^{(k)}_{n}$. These satisfy the same recursion relation given in Eq.\ref{eq:recur} for the $P^{(k)}_{n}$, but the initial conditions are different, $Q^{(0)}_{n}=Q^{(1)}_{n}=1$. The $P$ and $Q$ series are related by
\begin{eqnarray}
   Q^{(k)}_{n} = P^{(k)}_{n} + P^{(k-1)}_{n}
   \label{eq:relationpq}
\end{eqnarray}
The $P$ and $Q$ series are distinct, except for the Fibonacci case $n=1$. The $Q^{(k)}_{n}$ correspond to the number of letters in the $k$th approximant chain of the $n$th quasicrystal.

To build the 2D models of the next section, we use an alternative method of generating the metallic mean chains by the cut-and-project method starting from the 2D square lattice. The method has been described in detail elsewhere (see for example \cite{baake}), so we will be very brief. Consider a given value of $n$. To obtain the $k$th approximant, an infinite strip of rational slope $P^{(k-1)}_{n}/P^{(k)}_{n}$ is drawn, such that the width of the strip corresponds to a unit cell of the lattice. The approximant chain is obtained by projecting the points lying within the strip onto a line. The nearest neighbor distances along the resulting 1D chain are A (long) or B (short), and occur in a sequence which repeats periodically. The unit cell comprises $Q^{(k)}_{n}$ sites. We note that choosing different positions for the selection strip does not result in a new approximant for an infinite sample. In contrast, for a finite sample, shifting the selection strip results in different finite chains. This fact is relevant later on, when we consider edge states for systems with open boundary conditions.

\section{Tight-binding Hamiltonians in 1D and 2D and topological invariants}
We consider hopping (or off-diagonal) Hamiltonians on approximants of the metallic chain of degree $n$ and generation $k$ having the form
\begin{eqnarray}
    H^{1D}(n,k) = \sum_i t_i (c^\dag_i c_{i+1} + h.c.)
    \label{eq:hamilt}
\end{eqnarray}
where the summation is over site indices $i=1,...,N$ where $N$ is the number of sites. Hopping amplitudes can take the values $t_A$ or $t_B$, according to the sequence of A and B in a given approximant ${\mathcal{C}^{(k)}_n}$. The spectrum of the Hamiltonian in Eq.\ref{eq:hamilt} is symmetric around $E=0$ due to the chiral symmetry present in the model. Without loss of generality, both hoppings will be considered to be positive, since their signs can be gauged away. This family of models, extensively studied for $n=1$, has also been investigated for other values of $n$. As in the golden mean chains, the band structures in the silver mean chains are hierarchically related (see \cite{rudinger1998multifractal} as the index $k$ is increased. A perturbative renormalization scheme, analogous to the theory for $n=1$, has been introduced in \cite{thiem2009, thiem2011,thiem2012,rolof2013} although this approach has thus far not led to explicit solutions for spectral bands and gaps. Analogously to the well-known $n=1$ model, electronic spectra for $n>1$ should also present self-similar behaviors, and this is indeed observed. Even if the theoretical challenge of finding recursion relations remains, there are reasonable grounds to expect that all the metallic mean chain approximants have fractal band structures, and also that all of their possible gaps are open as long as $t_A \neq t_B$. In the Supplementary Information (SI) we present numerical data which illustrate that band structures are hierarchically constituted -- spectra and gaps of larger chains can be related to those of smaller chains. They are given for small values of $n$ and $k$ (see Fig.1 of the SI), but should suffice to give the overall picture. Gaps of all sizes are present, and it is checked numerically (see Fig.2 of the SI) that gap widths converge to finite values as system size increases. 
To compute topological numbers, we will proceed according to the scheme given in \cite{jaga2025}, linking the 1D Fibonacci chain to a 2D Quantum Hall problem. The 2D model describes electrons hopping on a square lattice submitted to a fictitious ``geometric flux" whose value is imposed by the slope of the selection strip. For our metallic mean approximants, this geometric flux is chosen to be
\begin{eqnarray}
    \phi^{(k)}_{n} = \frac{P^{(k)}_{n}}{Q^{(k)}_{n}},
\end{eqnarray}
in units of the flux quantum $\phi_0$. Changing $\phi_n^{(k)}$ to $1- \phi_n^{(k)}$ amounts to replacing the slope of the selection strip by its reciprocal, and results in all of the A bonds being replaced by B bonds and vice-versa. Just as in the original QH problem \cite{hofstadter}, it is convenient to work in the Landau gauge and take the fictitious vector potential parallel to the $y$ axis. The 2D quantum Hall problem is then described by the Hamiltonian 
\begin{eqnarray}
    H^{2D}(\phi^{(k)}_{n}) = \sum_{l,m} (t^{y}_{l,m} c^\dag_{l,m}c_{l,m+1} + h.c.)   + t^{x}_{lm} (c^\dag_{l,m}c_{l+1,m} + h.c.) 
    \label{eq:fhham}
\end{eqnarray}
where sites are labeled according to their positions $\vec{r}_{lm}= la\vec{x} + ma\vec{x} $, where $a$ is the lattice constant. The hopping amplitudes along vertical bonds $t^{y}_{lm}= t_B \exp(2\pi i l \phi^{(k)}_{n})$ depend on the position, while the hopping amplitudes along horizontal bonds $t^x_{lm}=t_A$ are uniform.  In the case of isotropic hopping, $t_B=t_A=t$ and in the general case below, the hopping amplitudes will also henceforth be expressed in units of $t$.  

Arguments presented in \cite{jaga2025} for the golden mean quasicrystal, and which we will not reproduce here, show that there is adiabatic continuity going from the isotropic 2D QH model to the 1D quasiperiodically modulated 1D system. The passage from 2D to 1D is achieved in two stages: i) starting with the isotropic 2D QH problem one  progressively transforms the system into a set of parallel chains of approximants, and ii) then increasing $t_B-t_A$ from 0 to its final value, during which process all gaps are expected to be opened. We recall that adiabatic continuity does not hold if the dimensional reduction from 2D to 1D is carried out by directly setting $t_A\neq t_B$ from the start since topological transitions will occur (as one can show by computations of topological markers using a spectral localizer formalism \cite{cerjan}). The link between the 2D and 1D problems means that topological invariants obtained for the 2DQH model carry over to the 1D metallic mean chains by adiabatic continuity. Below, our computations of topological invariants will be carried out in the 2D model. 

Our labeling scheme for the gaps of the $k$th approximant of the $n$th metallic mean chain can now be given. The integrated density of states (IDOS) in each of the gaps, $I_{n,k}(j)$, equal to the fraction of states lying below the $j$th gap, can be written as 
\begin{eqnarray}
    I_{n,k}(j) = p_j + q_j \frac{P^{(k)}_{n}}{Q^{(k)}_{n}}
    \label{eq:indexing}
\end{eqnarray}
where $p_j$ and $q_j$ are integers called gap labels, with $\vert q_j\vert \leq N/2$. This Diophantine equation Eq.\ref{eq:indexing} is one of the principal results of the paper, and can be solved to determine the set of $p_j$ and $q_j$ for each of the gaps, $j=1,...,N-1$. The index $q_j$ is sufficient to classify a gap since the index $p_j$ is then fixed by the condition $0\leq I \leq 1$. 
Since indexation in finite systems is not unique, we will present numerical evidence below that Eq.\ref{eq:indexing} correctly gives the edge state winding numbers.

\begin{figure*}[h!] 
\includegraphics[width=0.2\textwidth]{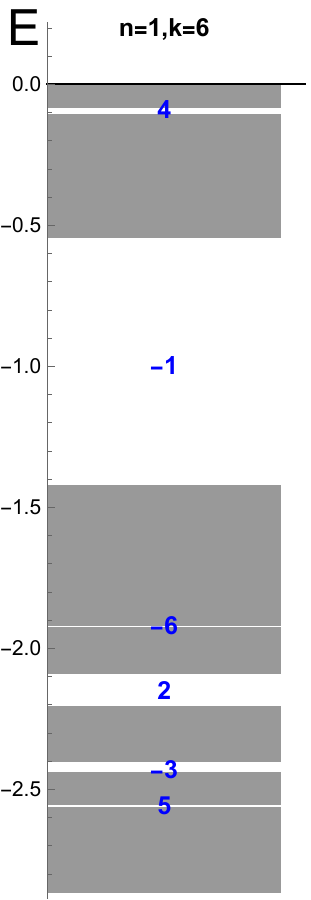}  
\includegraphics[width=0.2\textwidth]{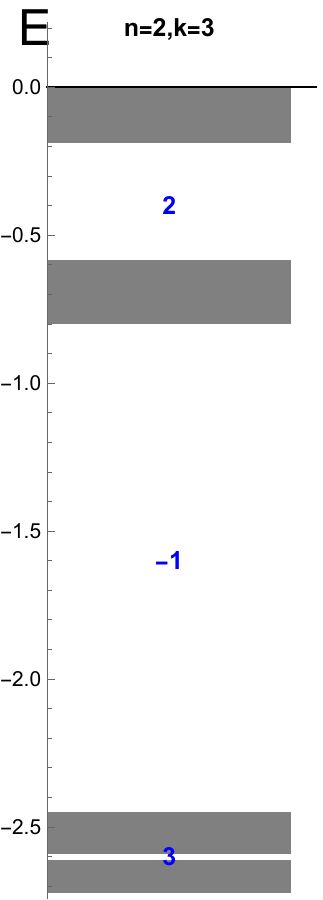} 
\includegraphics[width=0.2\textwidth]{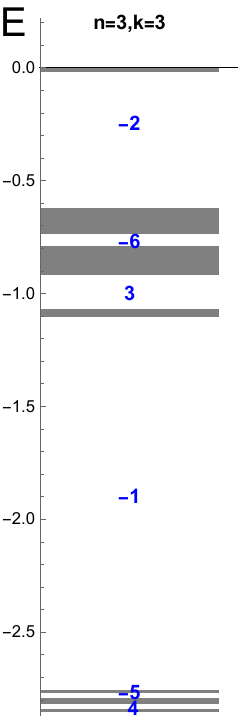}
\caption{The energy spectra $E$ of the first three metallic mean chains, for small k (values of $n$ and $k$ as indicated in each figure). Bands are shown in grey, gaps in white. The absolute values of the gap index are indicated in blue within each of the gaps. For reasons of symmetry only the lower half of the spectrum is shown. }
\label{fig:spectra}
\end{figure*}

In Fig.\ref{fig:spectra} energy spectra for the 2D model are plotted for the first three values of $n$ corresponding to gold, silver and bronze means. For clarity, the approximants shown are small enough, so that bands and gaps can be distinguished. The flux values are  $\frac{8}{13}$, $\frac{5}{7}$ and $\frac{10}{13}$. The values of gap labels are indicated in each of the gaps.
As the figures show, gap labels do not follow a simple ordering for small $n$ (we will see below a simplification for $n$ large). 
For the $n=1$ case, the $q=\pm 1$ gap is largest for all of the approximants. It has been observed, for the Fibonacci model, that gap widths tend to decrease with increasing index $q$, although the behavior is not monotonic \cite{maceicq}. When moving up the hierarchy of approximants,  the gap labels stabilize increases. The IDOS in any given gap tends to a fixed value as $k$ is increased, with 
\begin{eqnarray}
   I_n(q) = \lim_{k\rightarrow\infty} I_{n,k} (j)  \\ \nonumber
     I_n(q) = \mathrm{Mod}[q \frac{\omega_n}{1+\omega_n},1]
     \label{eq:limitI}
\end{eqnarray}
That is, for $n$ fixed, $p$ and $q$ for a given gap do not change as the size of the approximant is increased. This statement holds for the gaps which are termed stable, while the so-called transient gaps disappear for larger $k$ \cite{maceicq}. The limiting case of the quasicrystal is thus seen, as expected, to be described by the gap labeling theorem \cite{bellissard}.


The bi-infinite set of indices $n$ and $k$, $k=1,...\infty$ for $n=1,...,\infty$, defines a subset of the fractal Hofstadter butterfly diagram. The labeling of the gaps of the spectra has a complex structure. However, as $n$ becomes very large, the relations Eq.\ref{eq:recur},\ref{eq:relationpq} and \ref{eq:indexing} together imply that the labeling becomes progressively simpler. Most of the energy levels group together so as to order the gap labels $q$. The fraction of levels not concerned by this grouping tends to zero as $\sim 1/n$. One thus obtains an indexation of the gaps that increases systematically in absolute value -- 1,2,3,...starting from the band bottom. This is the expected small field behavior in the Hofstadter problem, where one finds Landau levels which are exactly equidistant, equidegenerate, and gap labeling of gaps occurs in a consecutive way. This last feature is seen in the purely 1D problem without any magnetic flux, as will be discussed in the following section devoted to the 1D butterfly diagram.

Concluding this section, we mention diagonal quasiperiodic Hamiltonians. In the simplest form, this can be a model in which the hopping is uniform along the approximant chain, whereas the onsite energies can take values $V_A$ and $V_B$. It is known that such a diagonal model is topologically equivalent to the 2D Quantum Hall problem \cite{zilberberg,zilberberg1}which is also the topological ``ancestor" of the off-diagonal models, as we have said. Thus, although we have focused here on the pure hopping ``off-diagonal" Hamiltonians, the gap labeling proposed here should apply quite generally to diagonal or mixed forms of these quasiperiodic Hamiltonians, provided that level crossings do not occur. An example of an adiabatic path linking diagonal to off-diagonal models is described in section 2 of the SI.


\section{A butterfly diagram for strictly 1D models}
The structure of the spectra of the 1D metallic mean quasicrystals is topologically equivalent to that of a subset of the Quantum Hall problem. It is interesting to see, furthermore, that a Hofstadter butterfly-like diagram is obtained upon plotting the spectra versus $\Omega_n = \omega_n/(1+\omega_n)$. To avoid clutter we do not represent the infinite set of approximants, though they can also be added. In principle the spectrum for each value of $n$ is a Cantor set, however, numerically, one can see the principal gap structures in the butterfly by considering sufficiently large approximants. Fig.2a) has been plotted for parameters $t_a=1$, $t_b=0.35$  while Fig.2b) has been plotted for parameters $t_a=1$,$t_b=2.5$. The values of n vary from 1 to 50. Periodic boundary conditions were imposed. The upper and lower band edges have been outlined.  

The main gaps of the spectrum, are shown outlined in red (for $q=\pm 1$), blue (corresponding to $q=\pm 2$) and green (corresponding to $q=\pm 3$). As $n$ is increased, one sees that the ordering of gap labels is changed -- for example, the gap $q=2$ goes above the gap $q=-1$ when going from the golden to the silver mean chain. It is easy to find the values of $n$ for which crossing of labels occur: for example, for small $n$, gap label 2 lies above gap label 3 (as in our examples shown in Figs.2), but their places are switched for $n$ large enough, and such that
\begin{eqnarray}
  \mathrm{min}[I_n(q=2),I_n(q=-2)]< \mathrm{min}[ I_{n}(q=3),I_{n}(q=-3)]
\end{eqnarray}
Using the relation for $I$ of Eq.\ref{eq:limitI} this gives the ``label-crossing" to occur between $n=3$ and $n=4$.  Other crossings can be similarly located.

As mentioned in the previous section, the ordering of the labels becomes extremely simple for $n\rightarrow \infty$. One sees at the right bottom corner of Fig.2a, how some of the principal gaps corresponding to $q=1,2,...$ rise upwards as a function of decreasing $\Omega_n$. The increase becomes linear 
asymptotically (that is, for large values of n). Recall that the Landau level formula for electrons in 2D gives the $q$th level to be $E_q=(q + \frac{1}{2})\hbar \omega_c$. Here $\omega_c$ is the cyclotron frequency, which is proportional to the magnetic flux and inversely proportional to the effective mass.

In the 1D systems too, the “geometric flux” due to quasiperiodicity can be considered as giving rise to “Landau levels”. By fitting the slopes for different values of the hopping parameters, the ``effective mass" in the metallic mean quasicrystal family can be estimated. We find (for details see Sec.3 of the SI) that, when $t_B < t_A$, the effective mass is close to that for a periodic crystal with hopping amplitude of $t_A$. This is expected, given that the quasicrystal is composed of periodic units with hopping amplitude $t_A$ of length $n$ and $n+1$ which are linked by the weaker $t_B$ bonds. In contrast, for $t_A < t_B$ the effective masses are large for small $t_A$, and decrease towards the periodic value as $t_A$ approaches $t_B$. This can be understood as being due to the fact that the lowest bands become extremely narrow in this system, for small values of $t_A$. 

\begin{figure*}[h!] 
\includegraphics[width=0.45\textwidth]{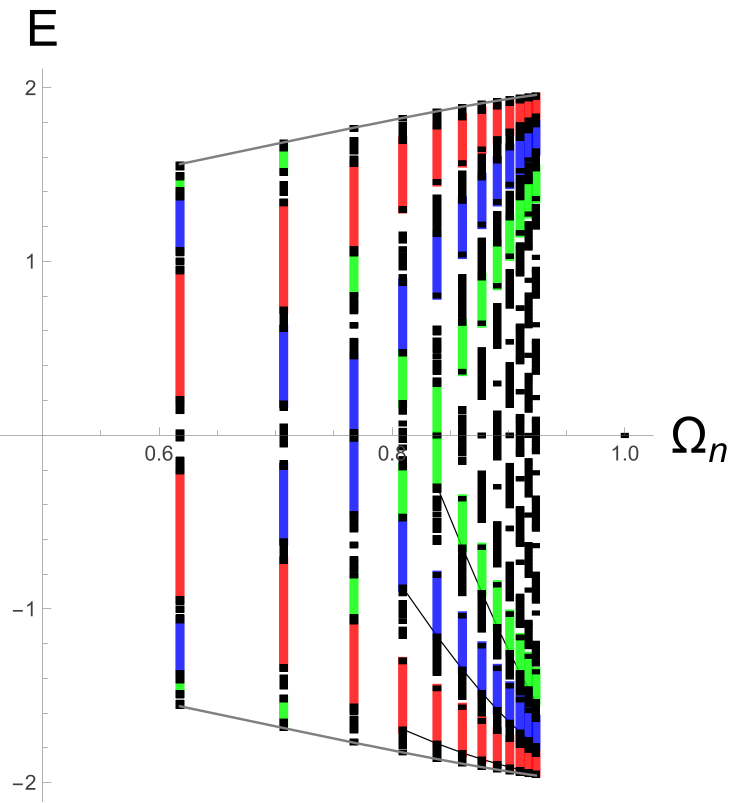}
\includegraphics[width=0.45\textwidth]{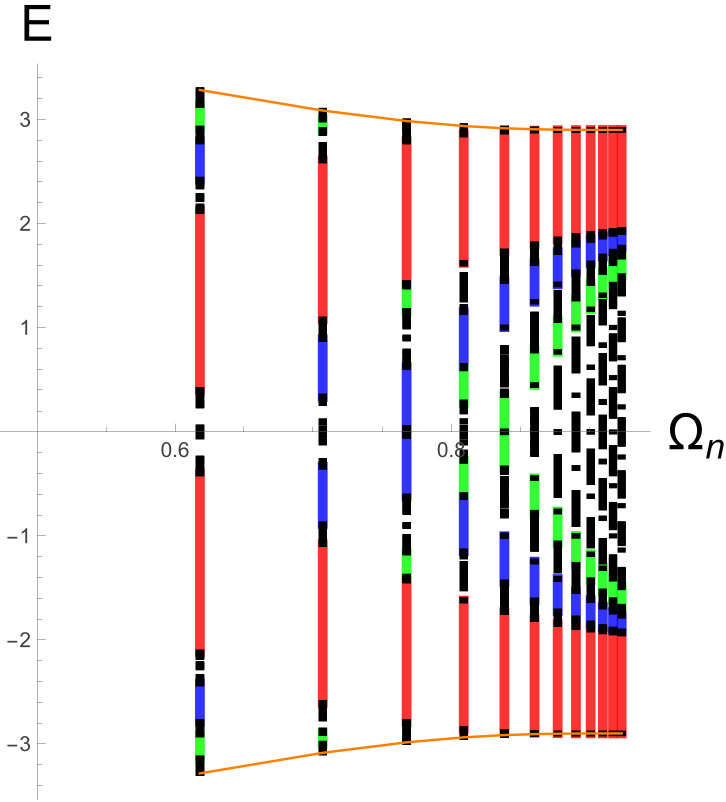} 
\caption{ Two Hofstadter “half butterfly” diagrams for the 1D hopping models of the metallic mean chains, Eq. (8). The energy spectra for several metallic mean chains (between n = 1 and n = 50) are plotted against $\Omega_n$ (see text for definition). Each of the individual spectra (series of vertical dots) are Cantor sets in the limit $k\rightarrow \infty$, but for the figures, they were numerically computed for large approximants, with $t_A$ = 1, $t_B$ =0.35 (Fig.2a) and with $t_A$ = 1, $t_B$ =2.5 (Fig.2b), under periodic boundary conditions. The outer band edges are joined as a guide to the eye (gray line). Intervals corresponding to the three lowest gap labels with |q|=1, 2, and 3 have been colored red, blue and green respectively in order to show how the ordering of color changes with increasing metallicity index n. In Fig.2a) the lines in the lower right hand corner tend asymptotically towards a linear dependence analogous to Landau levels in a 2D problem (see text). }
\label{fig:butterfly}
\end{figure*}


\begin{figure*}[h!]
\includegraphics[width=0.45\textwidth]{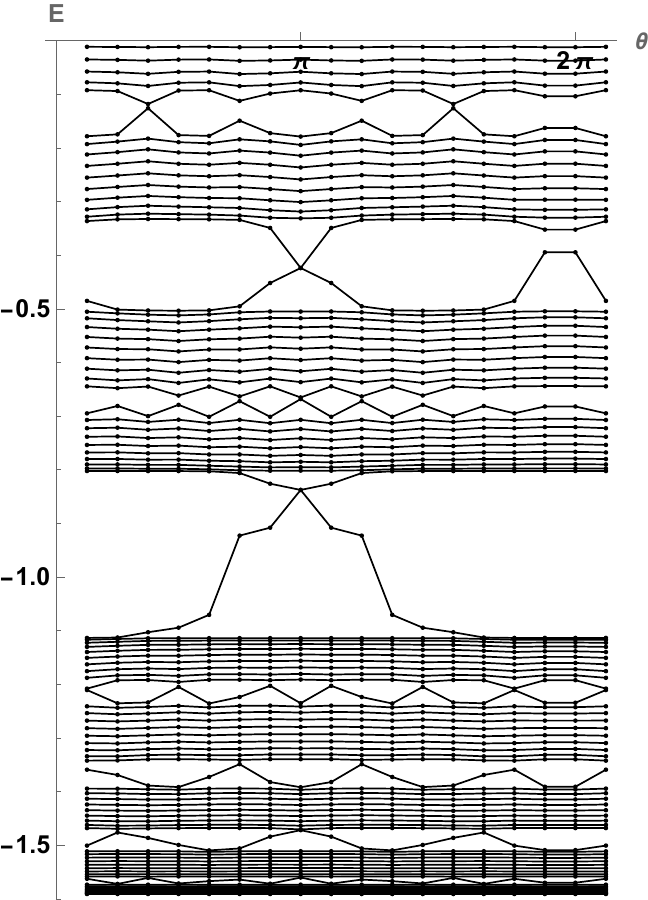} \hskip 1cm
\includegraphics[width=0.45\textwidth]{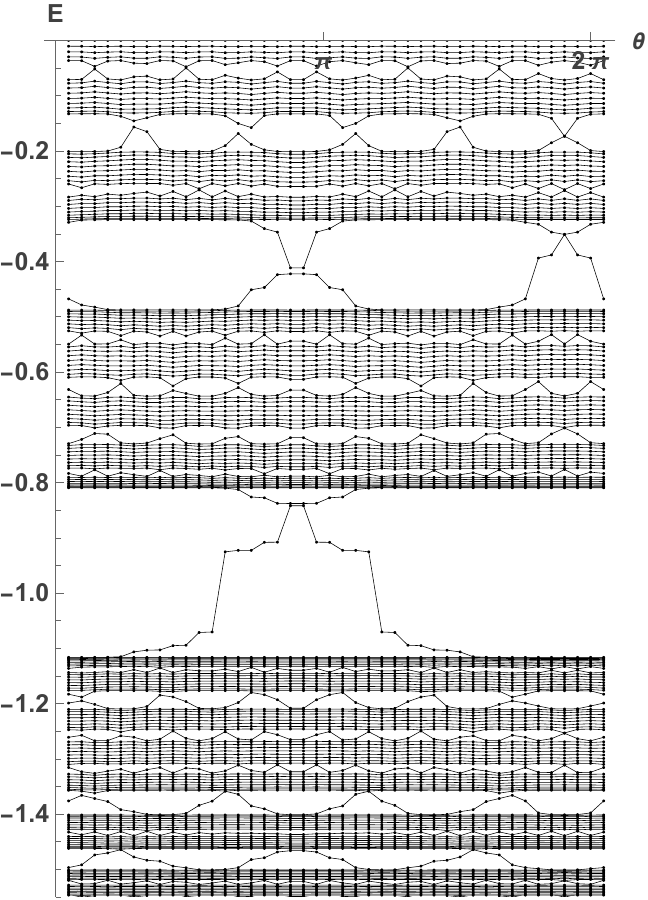} 
\caption{(left) Plot of the lower half of the energy spectrum of an open 1D $k=4$ silver mean approximant system of N=17 bands, showing the evolution of energies as the phason angle $\theta$ takes 17 discrete values between 0 and $2\pi$. A finite system of 8 unit cells was considered, with $t_A=0.7t,t_B=t$. The energies E is given in units of $t$. The lines are a only a guide to the eye, to help see the distinction between bulk states and edge states. The number of crossings of the energy of edge states in each gap can be seen to correspond to the gap label $q$ for this $n=2, k=4$ approximant chain. (right) Similar plot showing the winding of energies of the edge states in each of the gaps for an open 1D $k=5$ silver mean approximant system of N=41 bands. Note how the sequence of winding numbers in the gaps of the $k=4$ system (left hand figure) is reproduced in the larger $k=5$ system. }
\label{fig:phasonfig}
\end{figure*}



\section{Edge state winding in 1D metallic mean chains}
As mentioned in the previous section, the gap labeling scheme for finite chains in Eq.\ref{eq:indexing} will be checked by inspecting the behavior of edge states in open chains. According to the bulk-edge correspondence principle, one should expect topological edge states. These must exhibit a winding property of the edge states as a function of the so-called phason angle, $\theta$. The phason angle is a hidden degree of freedom of the quasicrystal system which is best visualized in the 2D representation as a displacement of the selection strip in the direction $y'$ perpendicular to the physical axis $x'$. For an approximant chain of $N$ sites, progressively shifting the selection window results in $N$ different chains, each one differing from the previous chain by one local permutation of bonds. The initial chain is recovered at the end of the cycle. During the phason cycle, the edge states in gaps shift in energy and position. As a function of the phason angle, edge modes can jump across the chain, giving rise to topological charge transport, giving a realization of a topological pump \cite{hatsugai2016}. These changes have been observed experimentally in photonic waveguides \cite{zilberberg} and in polaritonic resonators \cite{baboux}. The relation between winding of edge states and the topological invariants of the bulk, or bulk edge correspondence which was empirically observed in experiments has been given a theoretical foundation in \cite{prodan}.  

Fig.\ref{fig:phasonfig}a shows the energy bands and edge state energies in each of the main gaps as a function of the phason angle $\theta$ (for a definition of this angle see \cite{jagaICQ}) for a $k=4$ (17 site) silver mean approximant. Since the spectrum is symmetric, only the lower half of the spectrum is shown. Fig.\ref{fig:phasonfig}b shows similar data for the next largest approximant $k=5$ (41 sites). Considering the two figures enables one to visualize how the gap structures evolve as system size increases. In a given gap, the IDOS tends towards a fixed value as $k$ increases, and the gap topological indices remain invariant. The curves shown are not continuous, because the spectrum is only computed for a discrete number of values of $\theta$ (equal to the number of sites, 17), we have joined those data points. The ``oscillations" of the gap state energies are irregular, although one can see that they lead to the expected value of winding number $q$ for each gap. Indeed, one should note that an improved method by Kellendonk and Scaglioli uses a technique of ``augmentation" of the model that smooths the ``jumpy" behavior of edge state energies \cite{kellendonk} and it would be interesting to apply their method to this problem. 

We compare now the winding numbers deduced from Fig.\ref{fig:phasonfig} against the gap labels obtained using the indexing scheme in Eq.\ref{eq:indexing}. For our approximant chain, this scheme states that the $j$th gap labels are given by $j = 17p_j + 12q_j$ with $j=1,...,16$ such that $-N/2 \leq q_j \leq N/2$. The labels for the first 8 gaps $q_j$, starting from the bottom of the band are $q$= -7, 3, -4, 6, -1, -8, 2, and 5. These values correctly identify the number of periods of the motion of edge state in each of the gaps, as can be seen in Fig.\ref{fig:phasonfig}. Similar data is shown in Fig.3 of the SI for a bronze mean approximant chain, of $n=3,k=4$.

\section{Conclusions}
Topological invariants in quasicrystals are an active subject of research, with interesting potentials applications for topologically protected states. Studies have been carried out for topological quasiperiodic systems in one and also higher dimensions. However, those studies focused on the properties of specific structures, and a global picture of topological characteristics of quasicrystals has been lacking. In this work,
we have considered a family of 1D quasicrystals, and given a complete description of their topological invariants. These are the metallic mean chains, generalizations of the Fibonacci chain. We show how one can obtain the topological invariants for all of the approximants of all of the members of this family. This is done by considering their 2D parent Quantum Hall models. 

The gap labelings obtained with our scheme was checked by numerical computations in finite open chains. The $q$ indices correctly reflect the winding property of the edge states. This property has many implications for experimental systems. Edge states and their winding are interesting in many different contexts. One situation of interest occurs in models where one can induce superconducting pairing either by proximity effects \cite{gautam} or electron-electron interactions \cite{annica}. Majorana edge states \cite{annica} and Aharanov bound states \cite{josephson} with nontrivial winding properties have been predicted for the Fibonacci case and should also hold for the other members of the metallic mean family. 

We have shown that the spectra of metallic mean quasicrystals has a simplified Hofstadter butterfly structure. It is also interesting to note that ``Landau levels" inherited from the 2D system emerge in the asymptotic limit for the strictly 1D problem. These 1D Landau levels lie below the principal gaps of the spectrum, $q=1,2,3,...$, and correspond to the band fillings $q\Omega_n$. The effective mass obtained by fitting the slopes varies with the hopping parameters $t_A$ and $t_B$. A detailed study of the LL structure in the entire phase diagram remains for future work.

Other models sharing topological characteristics with the hopping model discussed here include, as we have mentioned, models with modulated onsite energies. More generally, the metallic mean family could provide interesting new scenarios for coupled and interacting systems. Coupling to phonons, for example, could lead to exotic new charge density modulated phases and topological phase transitions. Such models should be explored in future work. Finally, one can note that higher dimensional quasiperiodic structures based on metallic means have been proposed, as in the modulated honeycomb lattices in \cite{kogahc}. In these lattices, the structures tend towards the crystalline honeycomb structure asymptotically -- analogously to our 1D chains which tend towards the periodic chain. It would be interesting to understand the the variations of electronic properties of these lattices in that family of 2D structures, and their possible higher dimensional analogues.

\begin{acknowledgments}
I gratefully acknowledge numerous useful discussions with Frédéric Piéchon and Pavel Kalugin.
\end{acknowledgments}



\bibliography{mybib} 

\end{document}


\title{SI for Metallic mean quasicrystals and their topological invariants}

\author{Anuradha Jagannathan}
\affiliation{
Universit\'{e} Paris-Saclay, CNRS, Laboratoire de Physique des Solides, 91405 Orsay, France
}

\maketitle

\section{Convergence of spectra and gaps }
\begin{figure*}[h]
\includegraphics[width=0.8\textwidth]{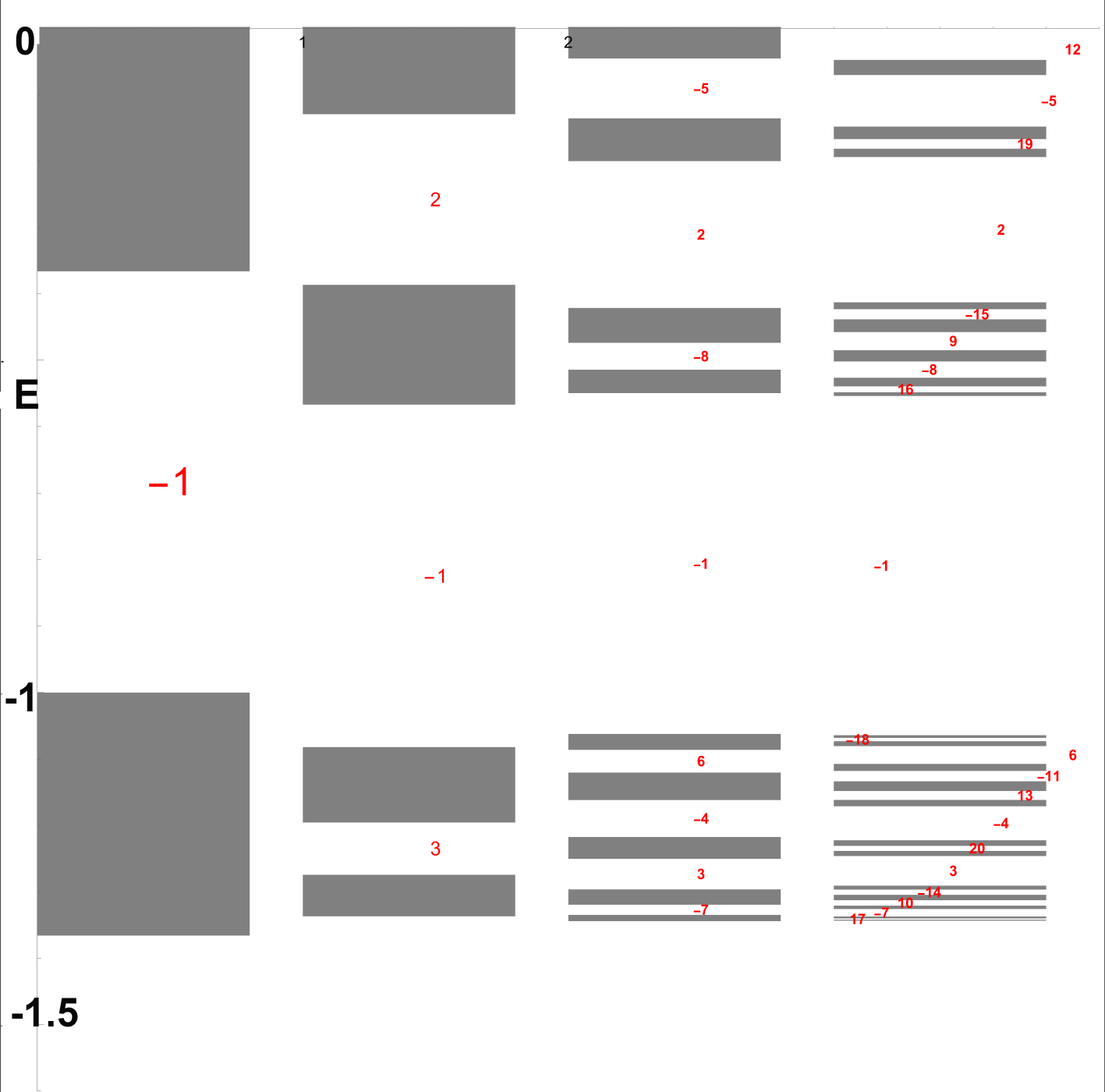}\hskip 2cm
\caption{The energy bands of four successive approximants of the silver mean quasicrystal ($n=2; k=2,3,4,5$ as computed numerically for $t_A=0.5, t_B=1$. The topological invariants $q$ of each of the gaps are indicated. Only the lower half of the (symmetric) spectra are shown. Comparing the spectra one sees how the gaps and their labels converge. }
 
\label{fig:sifig1}
\end{figure*}

In this section we will illustrate how the spectra and in particular, spectral gaps evolve as a function of the size of the approximant. When $t_A \neq t_B$, one expects that all of the $N-1$ gaps of the approximant composed of $N$ sites will be open. The distribution of gap widths has been computed explicitly for the $n=1$ golden mean approximants. The other metallic mean systems have been also studied, especially the cases of $n=2$, where a calculation using degenerate perturbation theory was shown to be in good accord with numerical spectra \cite{thiem2009}. For arbitrary values of the hopping ratio, an explicit non-perturbative construction of the hierarchical series of spectral gaps has been obtained in the case of $n=1$ \cite{rudinger1998multifractal}. No such analytical results have as yet been obtained for $n>1$. However, the recursive construction of the family of metallic mean chains implies that the gaps in large systems are derivable from those of smaller systems, and even if the gap widths can become very small there are no gap closings. As a consequence, they can be indexed as discussed in the main text with gap labels $q$ in the range $[-N/2,N/2]$. 
Fig.\ref{fig:sifig1} shows the gap structures for four consecutive silver mean approximant chains corresponding to $k=2 (N=3)$, $k=3(N=7)$, $k=4 (N=17)$ and $k=5 (N=41)$. As can be seen, the gap labels are stable when going from smaller to a larger approximant chain. For each value of the label $q$ (shown in red within each of the gaps) the IDOS can be seen to converge to well-defined values in the quasiperiodic limit. 

\begin{figure*}[h]
\includegraphics[width=0.5\textwidth]{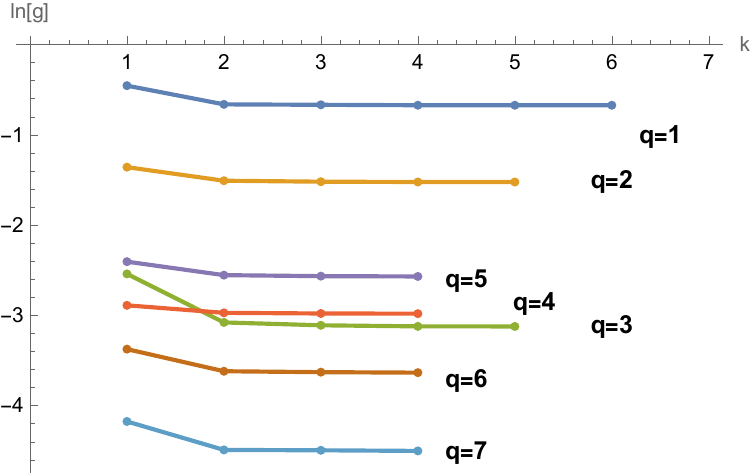}\hskip 2cm
\caption{ The logarithm of gapwidths for the seven largest gaps in the preceding spectra, plotted as a function of the index $k$ of the approximants. Each of the lines corresponds to a fixed value of $q$ ranging from 1 to 7. The curves are seen to flatten out to a constant non-zero value with increasing system size ($k$). The same is true for all of the other gapwidths, which are thus non-zero in the limit $k\rightarrow \infty$, as expected on theoretical grounds (see text).}
 
\label{fig:sifig2}
\end{figure*}

The variation of gap widths with system size, namely the value of $k$, for the 7 largest gaps, corresponding to labels $q=1,...,7$ is shown in Fig.\ref{fig:sifig2}. Each of the lines connects the gap widths of a fixed value of $q$. It can be seen that the convergence of the widths as $k$ increases is rather rapid for these gaps. We have checked that all of the smaller gaps converge as well. Note that the figure uses a logarithmic scale since the widths tend to decrease rapidly with increasing $q$.

\begin{figure*}[h]
\includegraphics[width=0.4\textwidth]{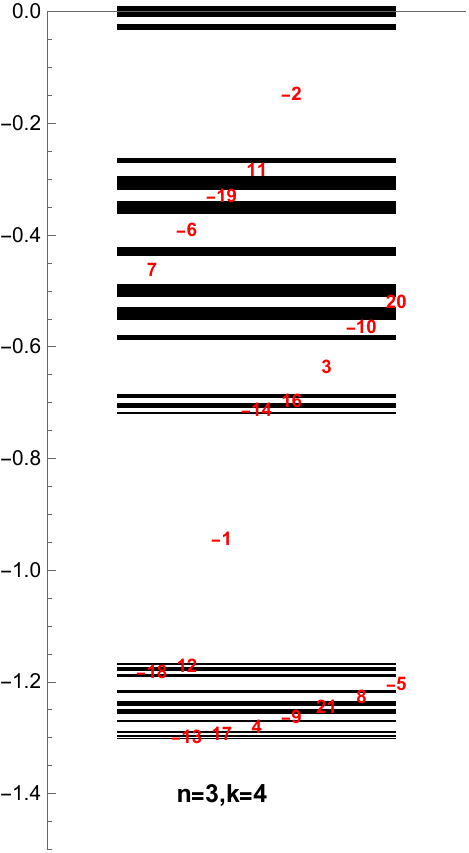}
\includegraphics[width=0.55\textwidth]{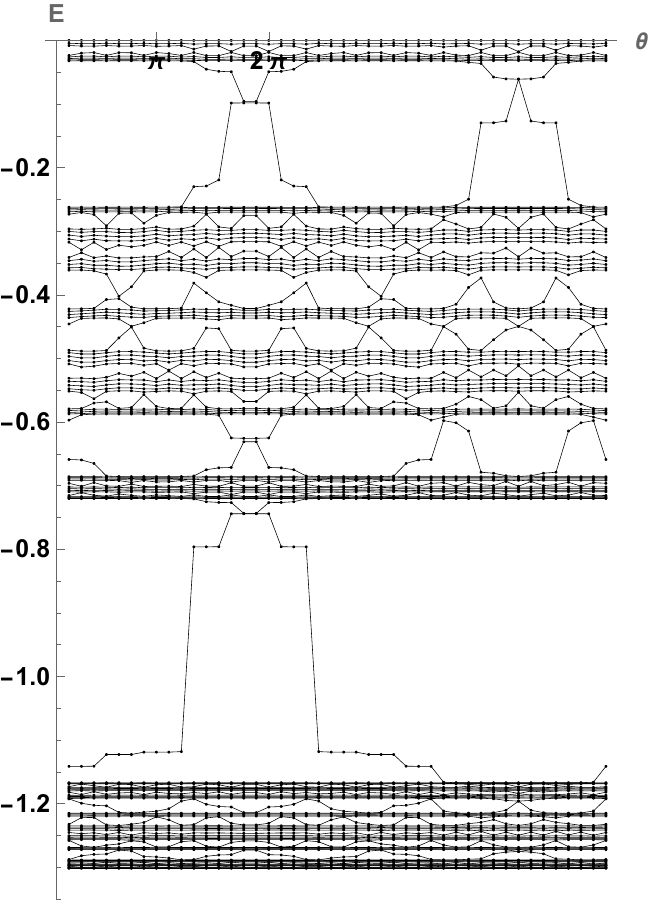}
\caption{(Left) The energy bands of an approximant of the bronze mean quasicrystal ($n=3; k=4$) as computed numerically for $t_A=0.5, t_B=1$. Only the lower half of the (symmetric) spectrum is shown. The topological invariants $q$ of some of the major gaps are indicated. (Right) Winding of edge states in an open $k=4$ bronze mean chain (43 sites) computed for the same hopping parameters. }
 
\label{fig:sifig3}
\end{figure*}

We show one more example of the spectrum for a 43 site approximant of the bronze quasicrystal ($n=3; k=4$). Fig.\ref{fig:sifig3} shows, on the left,the lower half of the energy bands of the approximant, computed numerically for $t_A=0.5, t_B=1$. The values of labels are shown for some of the gaps (the largest ones). The right hand figure shows the edge states which are present in each of the gaps when the chain is opened. Winding numbers can be counted and compared with the theoretically predicted values.
The winding is difficult to observe in the narrower gaps, but they can be seen better by taking larger system sizes.

\section{Adiabatic connection between off-diagonal and diagonal 1D quasicrystal Hamiltonians}
\begin{figure*}[h]
\includegraphics[width=0.5\textwidth]{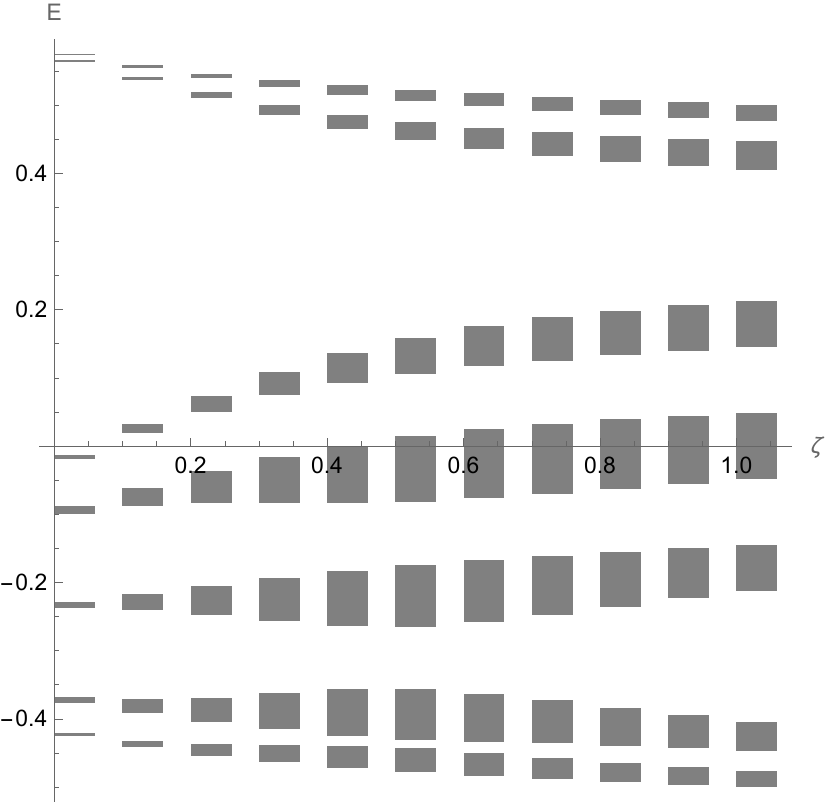}\hskip 2cm
\caption{The energy bands of the mixed Hamiltonian (Eq.1) for a 7-site silver mean approximant, computed numerically for $t_A=0.5$, $t_B=1$, $t_d=0.2$, $V_A = 2/7$ and $V_B =-5/7$ and shown for increasing values of $\zeta$ in the range [0,1]. For $\zeta=0$, one sees the spectrum of the diagonal model, while the spectrum of the off-diagonal model is attained for $\zeta=1$. Adiabatic continuity is preserved throughout the transformation.}
 
\label{fig:sifig4}
\end{figure*}

The off-diagonal Hamiltonians discussed in the main text (Eq.6) share topological characteristics with diagonal models. This can be shown by considering a mixed Hamiltonian, which depends on a continuous parameter. It combines the pure hopping Hamiltonian $H_{od}$ comprising two hopping amplitudes (as in Eq.6 of the main paper), and a purely onsite (diagonal) Hamiltonian $H_d=\sum t (c^\dag_i c_{i+1}+h.c.)+ \sum V_i c^\dag_i c_i$, where the hopping amplitude $t$ is uniform along the chain, while the onsite energies $V_i$ can take the values $V_A$ or $V_B$. We saw that for $H_{od}$, quasiperiodicity was controlled by the ratio of hopping amplitudes $t_A/t_B$ (note that their signs are unimportant as they can be gauged away). For the diagonal case the quasiperiodicity of $H_d$ is governed by the difference of the potentials $V_A/t$ and $V_B/t$. These can be normalized, and also shifted so as to correspond to an average value of zero. Consider the mixed form
\begin{eqnarray}
    H(\zeta) = \zeta H_{od} + (1-\zeta) H_d
\end{eqnarray}
where the parameter $\zeta$ allows one to go from the purely diagonal to the purely off-diagonal case. 
We consider $H(\zeta)$ for an approximant corresponding to some given $n$ and $k$ and how the spectral gaps evolve with $\zeta$. The spectrum for $\zeta=0$ is that of the diagonal model $H_d$. For $\zeta=1$, the spectrum is exactly symmetric, corresponding to the purely off-diagonal model. There are no band crossings, thus there is adiabatic continuity and topological numbers remain the same while going from diagonal to off-diagonal models. The spectra obtained by numerical computations for a  silver mean approximant of $7$ sites is shown in Fig.\ref{fig:sifig4}. Without loss of generality, one can choose the potentials such that their average value is zero. A typical choice for parameters is taken as $t_A=0.5$, $t_B=1$, $V_A = 2/7$ and $V_B =-5/7$. Note that, for better visualization, all the spectra have been normalized so as to conserve a constant band width of unity, for each value of $\zeta$.  

 \section{Landau levels in the 1D spectra in parameter space}
This section presents some details about spectra of 1D metallic mean chains and some results of fits to extract slopes of the "Landau levels" observed at the corners of the butterfly diagram shown in Fig.2 of the main text).
In our numerical work, we consider the $n$th metallic mean chain, and take a large enough approximant such that the principal bands and gaps can be considered to have converged sufficiently to their asymptotic value. The total number of bands is $Q^{(k)}_n$. The $q$th ``LL"  is the energy level of index $q \Omega_n$ where $q=1,2,...$. Plotting the energies for fixed q, as a function of $\Omega_n$, one obtains a set of straight lines, as can be seen in the butterfly figure. Recall for the 2D electron in a magnetic field that the standard formula for LL energies is $E_q = (q + \frac{1}{2}) \hbar\omega_c$, where $\omega_c$ is the cyclotron frequency. We recall that the cyclotron frequency depends on the magnetic field and the effective mass as $\omega_c = eB/2m_{eff}$. It is easy to show that the 2D lattice energy levels scale with flux as $E_q = \alpha (q + \frac{1}{2}) \Omega_n$, where $\alpha$ is inversely proportional to the effective mass $m_{eff}$. In particular, the value of $\alpha = 2\pi t_A$ for the periodic crystal where all hoppings are equal to $t_A$. 

In our 1D quasicrystal family, we have assumed that LL levels follow the form
\begin{eqnarray}
    E_q (n;k) = \alpha (t_A,t_B) (q + \frac{1}{2} + c_1) \Omega_n + c_2
\end{eqnarray}
where $\alpha$ depends on the hopping parameters. The additional constants have been introduced to account for the change of energy spectrum in going from the 2D to the 1D model.
Fig. 5 shows the data obtained by fitting the first five LL for a variety of different values of the hopping parameters $t_A$ and $t_B$ and approximants upto $n=15$. For $t_A=t_B$, gaps are closed and the LL are not defined. It can be seen that the red data corresponding to $t_B < t_A$ have a rather flat dependence on $t_B$ and lie close to the expected value for the periodic crystal. This is easily understood in the limit of small $t_B$, when the spectra can be computed in perturbation theory. In contrast, the blue data corresponding to $t_A < t_B$ show that $\alpha$ increases linearly with $t_A$ reaching the periodic crystal value when $t_A \approx t_B$. In other words, the effective mass is decreasing from high values for small $t_A$. This is expected, since the energy bands at the bottom of the spectrum are extremely narrow in this limit, where it is necessary to use perturbation theory to very high order in $r$. It will be interesting to explain this behavior with the help of a complete theory, in future work. 

\begin{figure*}[h]
\includegraphics[width=0.5\textwidth]{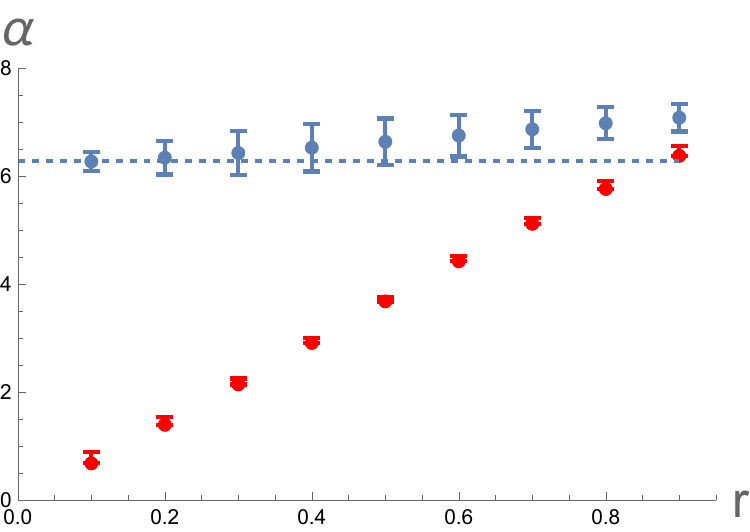}\hskip 2cm
\caption{The values of the LL slope factor $\alpha$ in parameter space. The red data with error bars show values of $\alpha$ for different values of $r=t_B/t_A$. The blue data shows values of $\alpha$ for different values of $r=t_A/t_B$. The dashed line represents the value $\alpha=2\pi$ expected for a periodic chain of hopping amplitude $t_A=1$. }
 
\label{fig:sifig5}
\end{figure*}

\bibliography{mybib} 